\documentclass[twocolumn]{ceurart}

\usepackage[frozencache]{minted}
\setminted{breaklines=true}

\begin{document}

\copyrightyear{2023}
\copyrightclause{
    Copyright for this paper by its authors.
    Use permitted under Creative Commons License Attribution 4.0 
    International (CC BY 4.0).
}

\conference{
    ITAT'23: Information technologies -- Applications and Theory
    September 22--26, 2023, Tatranské Matliare, Vysoké Tatry
}

\title{Prefix-free graphs and suffix array construction in sublinear space}

\author[1]{Andrej Baláž}[
    email=andrejbalaz001@gmail.com,
]
\cormark[1]

\author[1]{Alessia Petescia}[]

\address[1]{
    Department of Applied Informatics,
    Faculty of Mathematics, Physics and Informatics,
    Comenius University, Bratislava, Slovakia
}

\cortext[1]{Corresponding author}

\begin{abstract}
A recent paradigm shift in bioinformatics from a single reference genome to a pangenome brought with it several graph structures.
These graph structures must implement operations, such as efficient construction from multiple genomes and read mapping.
Read mapping is a well-studied problem in sequential data, and, together with data structures such as suffix array and Burrows-Wheeler transform, allows for efficient computation.
Attempts to achieve comparatively high performance on graphs bring many complications since the common data structures on strings are not easily obtainable for graphs.
In this work, we introduce prefix-free graphs, a novel pangenomic data structure; we show how to construct them and how to use them to obtain well-known data structures from stringology in sublinear space, allowing for many efficient operations on pangenomes.
\end{abstract}


\maketitle

\section{Introduction}
The term pangenome was first used by \citet{tettelin2005genome} in 2005 while studying variations in the population of Streptococcus agalactiae.
Since then, pangenomes have found applications in the study of many organisms, from viruses \cite{lau2021profiling} through microbes \cite{dutilh2014comparative} and plants \cite{danilevicz2020plant} to humans \cite{wang2022human}.
As per the definition by The Computational Pan-Genomics Consortium \cite{computational2018computational}, a pangenome is any set of genomic sequences meant to be analyzed jointly.
Nevertheless, in practice, most pangenomes consist of genomic sequences of highly related organisms and therefore are highly repetitive.
Representation of this repetitive dataset by simple text is often inefficient and limits scaling in terms of algorithmic time and space complexity.
These limitations lead to the idea of representing pangenomes as graphs, where similar genomic regions are unified into nodes, and these nodes are connected to paths representing the original genomic sequences.

Several approaches to pangenomic graph construction exist, such as variation graphs \cite{church2015extending,garrison2018variation,garrison2023building}, cactus graphs \cite{paten2011cactus,hickey2023pangenome}, and Wheeler graphs \cite{gagie2017wheeler,2022pfwg}.
Most of these approaches require an initial local alignment of similar regions or a multiple sequence alignment, which makes them computationally expensive.
Here we present a new class of graphs, prefix-free graphs, which are orders of magnitude faster to construct.
Furthermore, we explore the connection between prefix-free graphs and suffix arrays.

A suffix array is a data structure from the stringology field with a massive impact on designing many efficient algorithms on strings.
Particularly in bioinformatics, it is responsible for the design of such data structures as the Burrows-Wheeler transform \cite{burrows1994block} and FM-index \cite{ferragina2000opportunistic}, which in turn allowed for efficient mapping of reads to the reference and several other fundamental bioinformatics operations.
These fundamental operations are well-studied on sequential data, but the recent paradigm shift of moving from a single reference genome to a graph pangenome made it even more complicated to apply the acquired knowledge from the stringology field to biological sequences.

Thanks to the link to suffix arrays, prefix-free graphs have great potential to draw from this extensive knowledge.
Using the suffix array from a prefix-free graph, we can obtain several stringological data structures which are not easily obtainable for graphs.
This feature of prefix-free graphs was demonstrated in several articles \cite{2019boucher,rossi2022moni,2022pfwg,2022maria}, where the authors implicitly used similar techniques as presented here.
We think that explicitly defining and framing the prefix-free graph as a standalone pangenomic data structure can bring several benefits:

\begin{itemize}
    \item reduction in the complexity of the presentation of several space-efficient algorithms
    \item support of theoretical research by clearly delimiting the relevant terms
    \item improved focus on the optimization of algorithms related to prefix-free graphs
    \item enabling bringing prefix-free graphs closer to the biological data
\end{itemize}

In this work, we define prefix-free graphs and show how they can be constructed from the pangenome in its textual representation.
Furthermore, we show in detail how prefix-free graphs can be used to generate the suffix array of a pangenome in sublinear space and linear time.
Finally, we implement the presented algorithms as two binaries for easy construction of prefix-free graphs from a set of sequences in FASTA format and from a pangenomic graph in GFA format.
Furthermore, we implement the rust library for working with prefix-free graphs.
This library contains an iterator, which can be directly used to generate the suffix array in sublinear space.

\section{Prefix-free graphs}
The idea of prefix-free graphs is inspired by a technique used in the tool rsync named Context-Triggered Piecewise Hashing (CTPH) \cite{kornblum2006identifying}.
CTPH uses a rolling hash to partition a string into substrings such that long repeated substrings are partitioned the same way.
These substrings are then hashed with a traditional hash function and stored as a string signature.
The signatures of several files are then compared to determine changes.

In prefix-free graphs, we partition each sequence of a given pangenome into \emph{segment}s.
These segments form nodes of the prefix-free graph, and their adjacencies in the original sequences constitute edges.
The sequences are represented as \emph{path}s in the graph.

The segments have two essential characteristics making them a good choice for nodes of a pangenomic graph.
Similarly to CTPH, long repeated sequences will be partitioned the same way.
Furthermore, no segment is a prefix of another, making a set of segments prefix-free.
The second characteristic is crucial for connecting prefix-free graphs and suffix arrays, as will be presented in the next section.

To create a prefix-free graph from a given pangenome $P$, we define a set of trigger words $T$, where each trigger word is a string of length $k$.
For this set $T$, we build an Aho-Corasick automaton \cite{aho1975efficient}.
Then, for each sequence in pangenome $P$, we append $k$ sentinel characters and iterate over such modified sequence, searching for matches with set $T$ using the automaton.
Each time we encounter a trigger word, we recognize a new segment from the start of the previous trigger word to the end of the current trigger word.
If the segment's sequence was not yet observed during the scan, we add it to the set of segments and assign a unique ID.
Each time we append the corresponding ID to the path representing the original sequence. 

Two special cases happen during the sequence scan, one at the beginning, when no previous trigger word was encountered, and another at the end, when the last $k$ characters are sentinels.
These are addressed by simply starting the first segment at the start of a sequence and ending the last at the sequence end.

\begin{figure}
    \centering
    \includegraphics[width=0.6\linewidth]{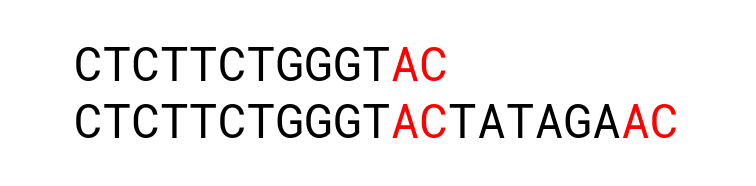}
    \caption{
        A sketch of a visual proof that the trigger words induce prefix-free segments.
        Suppose a segment is a prefix of another segment.
        By construction, it ends with a trigger word.
        This trigger word would be in the middle of another segment and would break it into two smaller segments.
    }
    \label{fig:proof}
\end{figure}

Notably, the adjacent segments overlap by exactly $k$ characters.
Furthermore, trigger words occur only at the beginnings or ends of segments because any occurrence of a trigger word in the middle of a segment would break it into two.
This feature and the choice of sentinels outside the sequence's alphabet guarantee that the set of segments is prefix-free.
A sketch of the proof is shown in Figure \ref{fig:proof}.

After the previous steps, the set of segments and the list of paths already represent a prefix-free graph.
However, to simplify the usage of prefix-free graphs, we recommend normalizing them.
During the normalization, we sort segments lexicographically and change their IDs to correspond to the lexicographical ranks.
This relabeling is then also propagated to paths accordingly. 

\begin{figure}
    \centering
    \includegraphics[width=\linewidth]{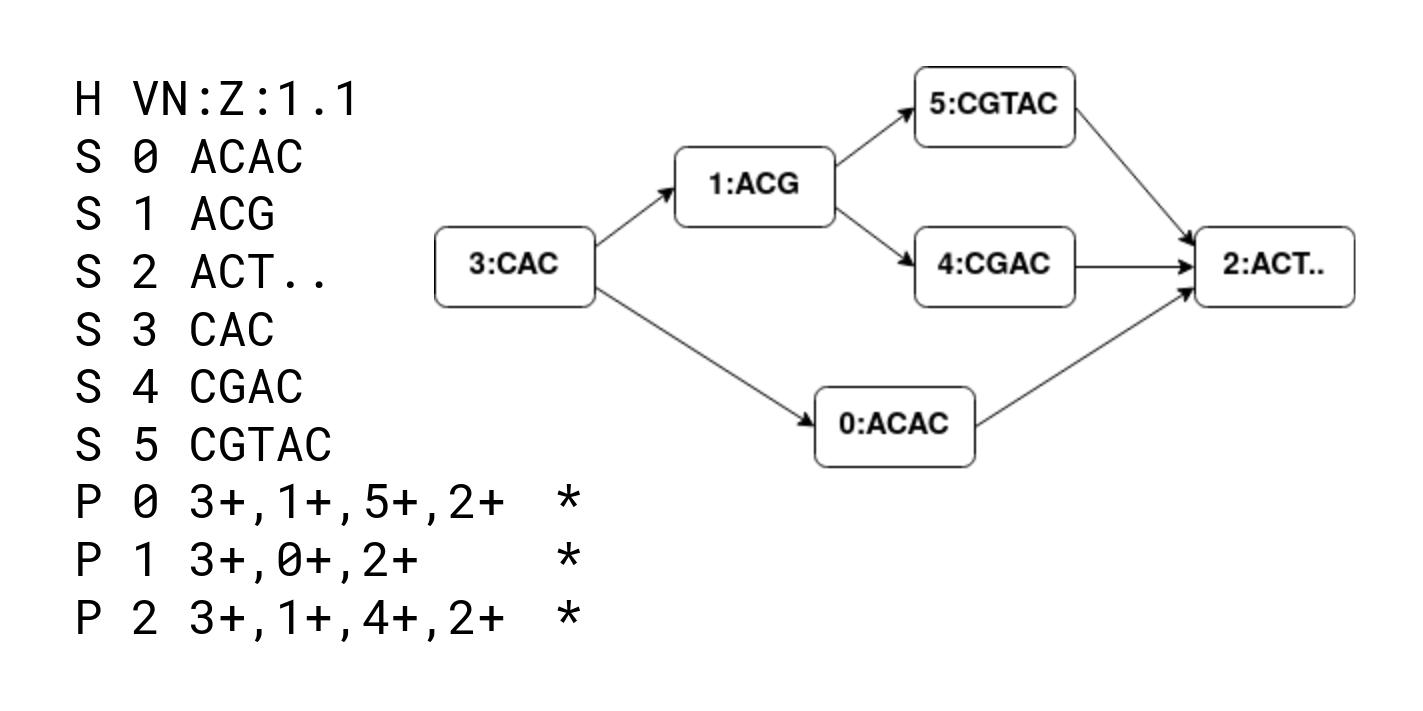}
    \caption{
        Prefix-free graph of the running example after normalization and its representation in GFA format.
        Link lines omitted for brevity.
    }
    \label{fig:gfa}
\end{figure}

To illustrate the entire procedure, consider a set of sequences $P = \{$\texttt{CACGTACT}, \texttt{CACACT}, \texttt{CACGACT}$\}$ and a set of trigger words $T = \{$\texttt{AC}, \texttt{CG}$\}$.
After the partitioning, we obtain a set of segments with IDs $\{$\texttt{0:CAC}, \texttt{1:ACG}, \texttt{2:CGTAC}, \texttt{3:ACT..}, \texttt{4:ACAC}, \texttt{5:CGAC}$\}$ and a list of paths \texttt{[[0,1,2,3], [0,4,3], [0,1,5,3]]}.
After the normalization, we get a prefix-free graph which can be directly represented in GFA format as shown in Figure \ref{fig:gfa}.

From this representation, original sequences of a pangenome can be reconstructed by expanding the segment IDs in a particular path, ignoring the last $k$ characters of each segment.

\section{Suffix array construction}
A suffix array is a permutation of string positions which lexicographically
sorts the suffixes of the string starting at that position.
It is an influential data structure with many applications in efficient string algorithms solving problems such as exact pattern matching, repeat finding, maximum exact match (MEM) finding, document retrieval and many more.

There exist several algorithms for suffix array construction in linear time \cite{skew2003karkkainen,sais2009nong,saoverview2007puglisi} with several practical implementations \cite{louza2017inducing,libdivsufsort}.
Despite their linear time complexity, these algorithms become bottlenecks in some applications because of their linear space complexity.
This observation is especially relevant in pangenomics, where the datasets often do not fit in the computer memory.

Here, we show another crucial advantage of prefix-free graphs.
Although they do not offer any improvement of theoretical guarantees in the worst case, in practice, they often represent the pangenome in a substantially smaller space and allow us to generate the suffix array values one by one, possibly using the values directly in subsequential computation or storing them in compressed form.
This iteration can be done without ever expanding the pangenome to its textual representation in space proportional to the sum of segment lengths and the sum of path lengths.

\subsection{Iterator preparation}
\begin{figure}
    \centering
    \includegraphics[width=\linewidth]{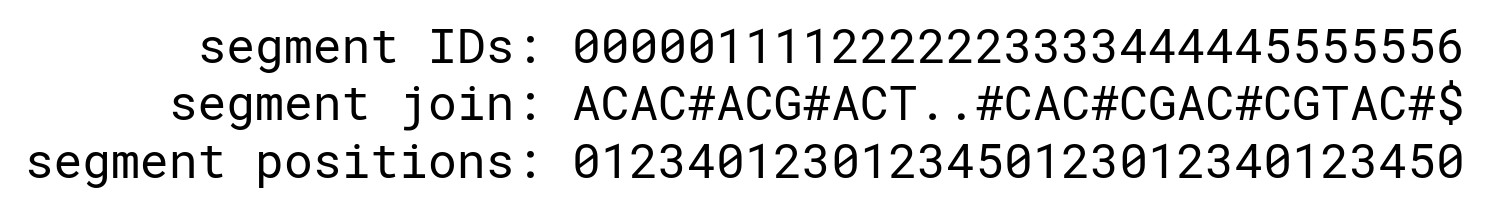}
    \caption{
        Segment join of segments from the running example, corresponding segment IDs, and positions.
        Permuting IDs and positions according to the inverse permutation of the suffix array results in the segment ID array and the segment position array shown in Table \ref{tab:suffix}.
    }
    \label{fig:ids_and_positions}
\end{figure}

To prepare the iterator of a suffix array of the pangenome from a prefix-free graph, we need to create several data structures.
First, we concatenate all the segments into a single string using a separator $\texttt{\#}$ and append a sentinel $\texttt{\$}$.
We will call this concatenation \emph{segment join}.
An example of a segment join is in Figure \ref{fig:ids_and_positions}.

Next, we calculate the segment join's suffix array and the longest common prefix (LCP) array.
For both of these arrays, there exist algorithms with linear time and space complexity which we can use.
We note that these linear complexities are proportional to the length of the segment join, which is usually much smaller than the original pangenome.

Next, for each suffix of the segment join, we need to calculate the corresponding segment ID and position values.
The value segment ID represents in what segment the current suffix starts, and the value segment position represents at what position in that particular segment the current suffix starts.
These arrays can be computed using an inverse permutation of a suffix array $\texttt{ISA}$ (Equation \ref{eq:isa}) of a segment join in linear time.

\begin{equation}
    \label{eq:isa}
    \texttt{ISA}[\texttt{SA}[i]] = i
\end{equation}

To illustrate the procedure, consider the segment join of our running example from Figure \ref{fig:ids_and_positions}.
Each position of the join can be assigned a segment ID and a position in the current segment by linearly scanning the segment join and incrementing the ID and position accordingly.
Then, applying the inverse permutation of a suffix array to these arrays changes the order of computed values in correspondence to the sorted suffixes.
The resulting suffix array, LCP array, segment ID array and segment positions array are stored in a \emph{suffix table} as shown in Table \ref{tab:suffix}.

\begin{table*}
    \caption{Suffix table}
    \label{tab:suffix}
    \begin{tabular}{rrrrrl}
        \toprule
        i & SA[i] & LCP[i] & ID[i] & pos[i] & suffix\\
        \midrule
         0  & 30 & -1 & 6 & 0 & \texttt{\textcolor{gray}{\$}}                                     \\
         1  & 29 &  0 & 5 & 5 & \texttt{\textcolor{gray}{\#\$}}                                   \\
         2  &  4 &  1 & 0 & 4 & \texttt{\textcolor{gray}{\#ACG\#ACT..\#CAC\#CGAC\#CGTAC\#\$}}     \\
         3  &  8 &  3 & 1 & 3 & \texttt{\textcolor{gray}{\#ACT..\#CAC\#CGAC\#CGTAC\#\$}}          \\
         4  & 14 &  1 & 2 & 5 & \texttt{\textcolor{gray}{\#CAC\#CGAC\#CGTAC\#\$}}                 \\
         5  & 18 &  2 & 3 & 3 & \texttt{\textcolor{gray}{\#CGAC\#CGTAC\#\$}}                      \\
         6  & 23 &  3 & 4 & 4 & \texttt{\textcolor{gray}{\#CGTAC\#\$}}                            \\
         7  & 13 &  0 & 2 & 4 & \texttt{.\textcolor{gray}{\#CAC\#CGAC\#CGTAC\#\$}}                \\
         8  & 12 &  1 & 2 & 3 & \texttt{..\textcolor{gray}{\#CAC\#CGAC\#CGTAC\#\$}}               \\
         9  & 27 &  0 & 5 & 3 & \texttt{AC\textcolor{gray}{\#\$}}                                 \\
        10  &  2 &  3 & 0 & 2 & \texttt{AC\textcolor{gray}{\#ACG\#ACT..\#CAC\#CGAC\#CGTAC\#\$}}   \\
        11  & 16 &  3 & 3 & 1 & \texttt{AC\textcolor{gray}{\#CGAC\#CGTAC\#\$}}                    \\
        12  & 21 &  5 & 4 & 2 & \texttt{AC\textcolor{gray}{\#CGTAC\#\$}}                          \\
        13  &  0 &  2 & 0 & 0 & \texttt{ACAC\textcolor{gray}{\#ACG\#ACT..\#CAC\#CGAC\#CGTAC\#\$}} \\
        14  &  5 &  2 & 1 & 0 & \texttt{ACG\textcolor{gray}{\#ACT..\#CAC\#CGAC\#CGTAC\#\$}}       \\
        15  &  9 &  2 & 2 & 0 & \texttt{ACT..\textcolor{gray}{\#CAC\#CGAC\#CGTAC\#\$}}            \\
        16  & 28 &  0 & 5 & 4 & \texttt{C\textcolor{gray}{\#\$}}                                  \\
        17  &  3 &  2 & 0 & 3 & \texttt{C\textcolor{gray}{\#ACG\#ACT..\#CAC\#CGAC\#CGTAC\#\$}}    \\
        18  & 17 &  2 & 3 & 2 & \texttt{C\textcolor{gray}{\#CGAC\#CGTAC\#\$}}                     \\
        19  & 22 &  4 & 4 & 3 & \texttt{C\textcolor{gray}{\#CGTAC\#\$}}                           \\
        20  &  1 &  1 & 0 & 1 & \texttt{CAC\textcolor{gray}{\#ACG\#ACT..\#CAC\#CGAC\#CGTAC\#\$}}  \\
        21  & 15 &  4 & 3 & 0 & \texttt{CAC\textcolor{gray}{\#CGAC\#CGTAC\#\$}}                   \\
        22  &  6 &  1 & 1 & 1 & \texttt{CG\textcolor{gray}{\#ACT..\#CAC\#CGAC\#CGTAC\#\$}}        \\
        23  & 19 &  2 & 4 & 0 & \texttt{CGAC\textcolor{gray}{\#CGTAC\#\$}}                        \\
        24  & 24 &  2 & 5 & 0 & \texttt{CGTAC\textcolor{gray}{\#\$}}                              \\
        25  & 10 &  1 & 2 & 1 & \texttt{CT..\textcolor{gray}{\#CAC\#CGAC\#CGTAC\#\$}}             \\
        26  &  7 &  0 & 1 & 2 & \texttt{G\textcolor{gray}{\#ACT..\#CAC\#CGAC\#CGTAC\#\$}}         \\
        27  & 20 &  1 & 4 & 1 & \texttt{GAC\textcolor{gray}{\#CGTAC\#\$}}                         \\
        28  & 25 &  1 & 5 & 1 & \texttt{GTAC\textcolor{gray}{\#\$}}                               \\
        29  & 11 &  0 & 2 & 2 & \texttt{T..\textcolor{gray}{\#CAC\#CGAC\#CGTAC\#\$}}              \\
        30  & 26 &  1 & 5 & 2 & \texttt{TAC\textcolor{gray}{\#\$}}                                \\
        \bottomrule
    \end{tabular}
\end{table*}

\begin{table}
    \caption{Segment table}
    \label{tab:segment}
    \begin{tabular}{rrll}
        \toprule
        id & length & starts & right context ranks \\
        \midrule
        0 & 4 & 9           & 9         \\
        1 & 3 & 15, 1       & 13, 14    \\
        2 & 5 & 18, 5, 11   & 1, 2, 3   \\
        3 & 3 & 8, 14, 0    & 4, 5, 6   \\
        4 & 4 & 16          & 7         \\
        5 & 5 & 2           & 8         \\
        \bottomrule
    \end{tabular}
\end{table}

\begin{figure}
    \centering
    \includegraphics[width=\linewidth]{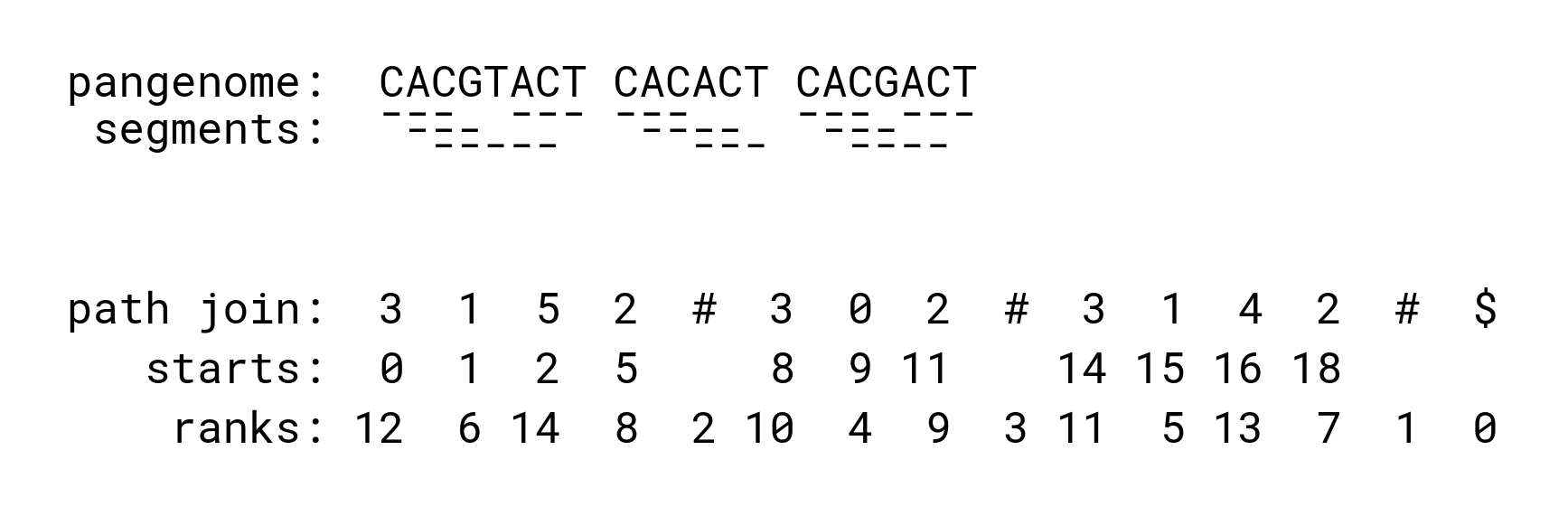}
    \caption{
        Path join of paths from the running example and the corresponding starts and ranks. The pangenome and segments are shown above to clarify the meaning of the start values.
    }
    \label{fig:path_join}
\end{figure}

In the suffix table, one row can represent multiple positions of the pangenome.
To identify these positions, we store some additional information in a \emph{segment table}.
For each segment, we store its length, starting positions in the pangenome and ranks of the right contexts of these positions.
To calculate the starting positions and the ranks of the right contexts, we use a \emph{path join}.
Similarly to segment join, a path join is a concatenation with delimiters $\texttt{\#}$ and sentinel $\texttt{\$}$, but now constructed by concatenating the paths.
An example of a path join for our running example is in Figure \ref{fig:path_join}.
Then, starting positions can be calculated by cumulatively summing the lengths
of the segments in path join and subtracting the overlaps.

The computation of ranks is more involved.
It uses the normalized form of prefix-free graphs since it relies on a lexicographically smaller ID in a path representing a lexicographically smaller segment.
We construct the suffix array of the path join and find its inverse permutation $\texttt{ISA}$.
$\texttt{ISA}$ gives us ranks for each position in the path join.
To determine the rank of the right context for position $i$, we take
the value of $\texttt{ISA}[i+1]$.
Finally, we store the starts and ranks sorted by the rank values in the segment table as shown in Table \ref{tab:segment}.

\subsection{Iteration}
With the previous tables stored in memory, we have all the necessary ingredients to generate the suffix array values.

Each row in the suffix table represents a suffix of a particular segment.
There are four cases of what the first position of these suffixes can represent within the segments:
\begin{itemize}
    \item the sentinel
    \item a separator
    \item a position within the last $k$ characters of a segment
    \item a position outside the last $k$ characters, separator and sentinel
\end{itemize}

Since the original pangenome has no corresponding position for the sentinel or separator characters, we can skip the first rows representing them.

In the third case, the position is inside the trigger word or the sentinels appended during the graph construction.
The positions inside the trigger words are represented twice in the suffix table, once at the end of a segment and a second time at the beginning of the following segment in the pangenome.
These ending positions can violate the prefix-free property of the segment suffixes and, therefore, can be sorted incorrectly.
Skipping through these positions ensures the prefix-free property for the rest of the suffixes and also avoids double reporting.
Therefore, if the length of a current segment suffix is smaller or equal to the size of the trigger words $k$, we skip the row as in the previous cases.
This choice also plays nicely with the previous choice of appending $k$ sentinels during the construction of a prefix-free graph, as these positions will not get reported either.

Finally, in the last case, we report the suffix array values.
The suffix table can be partitioned into blocks of the same segment suffixes.
For example, consider rows 20 and 21, which form a single block.
All other blocks in the running example consist of single rows; therefore, we call them singletons.

This partitioning leads to three cases:
\begin{itemize}
    \item a singleton block with segment suffix occurring only once in the whole pangenome
    \item a singleton block with segment suffix occurring several times in the pangenome
    \item a non-singleton block
\end{itemize}

In the first case, we must report only a single suffix array value.
Given the row index $i$, this value can be calculated with Equation \ref{eq:sa_value}.

\begin{equation}
    \label{eq:sa_value}
    \texttt{SA value} = \texttt{starts}[\texttt{ID}[i]] + \texttt{pos}[i]
\end{equation}

As an example, consider the row $13$ in Table \ref{tab:suffix}, the first row yielding a SA value.
Its segment ID is $0$, and from Table \ref{tab:segment}, we see only one occurrence of segment $0$ with starting position $9$ in the pangenome.
The offset from the start of a segment $\texttt{pos}[13]$ is $0$.
Summing these two values, we get the first value of a suffix array $9 + 0 = 9$ corresponding to the lexicographically smallest suffix $P[9..] = \texttt{ACACT}$.

The second, slightly more complex case is a singleton block representing a segment suffix with several occurrences in the pangenome.
In this case, we must report as many suffix array values as the number of occurrences.
Because the starting segment positions in the segment table are sorted based on their right context rank, we can iterate through these starting positions and apply Equation \ref{eq:sa_value} to each of them.

As an example, consider the row $14$ in Table \ref{tab:suffix}.
This suffix occurs twice in the pangenome in segments starting at positions $15$ and $1$.
Since the offset from the start of a segment $\texttt{pos}[14]$ is $0$, we report a suffix array values $15 + 0 = 15$ and $1 + 0 = 1$, corresponding to the suffixes $P[15..] = \texttt{ACGACT}$ and $P[1..] = \texttt{ACGTACT}$.

In the last case, we have a non-singleton block representing suffixes of several segments, possibly with multiple occurrences.
These suffixes represent identical substrings in the original pangenome.
Here, we report a suffix array value for each of the substrings.
To identify the first value, we must find the starting position with the smallest right context rank.
Because the ranks are sorted, this procedure is similar to the merging phase of a merge sort.
Therefore, to iterate through all suffix array values in the block, we always identify the segment start with the next smallest right context rank and apply Equation \ref{eq:sa_value} to this segment start.

As an example, consider the block of rows $[20..21]$ in Table \ref{tab:suffix}.
The relevant segment IDs are $0$ and $3$, with segment starts at positions $9$, $8$, $14$ and $0$.
The right context ranks from smallest to highest are $4, 5, 6, 9$ with the corresponding segment starts $8, 14, 0, 9$.
Applying Equation \ref{eq:sa_value} to these segment starts yields a suffix array values $8 + 0 = 8$, $14 + 0 = 14$, $0 + 0 = 0$ and $9 + 1 = 10$, representing pangenome suffixes $P[8..] = \texttt{CACACT}$, $P[14..] = \texttt{CACGACT}$, $P[0..] = \texttt{CACGTACT}$, and $P[10..] = \texttt{CACT}$.

\section{Results}
We implemented the prefix-free graphs as a Rust package.
This package contains two binary crates and one library crate.
Binary crates are executables that serve the purpose of creating prefix-free graphs from FASTA and GFA formats.
The binary crates are named \texttt{fasta2pfg} and \texttt{gfa2pfg}, and their usage is as follows:

\begin{minted}{bash}
fasta2pfg -t triggers.txt < pangenome.fna > pfg.gfa

gfa2pfg -t triggers.txt < pangenome.gfa > pfg.gfa
\end{minted}

We compared the running time of the construction algorithm with several pangenome construction tools, namely with the PanGenome Graph Builder \cite{garrison2023building}, VG \cite{garrison2018variation}, and Minigraph-Cactus \cite{hickey2023pangenome}.
For comparison, we used up to 256 SARS-CoV-2 sequences.
These sequences have lengths of around 29.9 kbp and high nucleotide similarity.
We used stop codons (\texttt{TAA}, \texttt{TAG}, and \texttt{TGA}) as trigger words to construct prefix-free graphs.
The resulting running times are shown in Table \ref{tab:comparison}.

\begin{table*}
    \label{tab:comparison}
    \begin{tabular}{rrrrr}
        \toprule
        \#seqs & pggb & vg & minigraph-cactus & pfg \\
        \midrule
        4   & 0m 12s & 0m 1s  & 3m 47s  & 0m 1s \\
        16  & 0m 19s & 0m 1s  & 6m 30s  & 0m 1s \\
        64  & 1m 20s & 0m 2s  & 13m 30s & 0m 1s \\
        256 & 7m 20s & 0m 55s & 41m 49s & 0m 3s \\
        \bottomrule
    \end{tabular}
    \caption{
        Comparison of running times of pangenomic graph construction tools.
        Wall clock times are measured by the built-in bash command \texttt{time} and rounded up to the nearest second.
        In addition to the measured times, \texttt{pggb} and \texttt{vg} need preprocessing of the pangenome.
        \texttt{pggb} requires the fasta file to be indexed.
        This time was negligible in our case, but it can have a more significant impact when a more extensive dataset is considered.
        \texttt{vg} builds a pangenome from a VCF file or a multiple sequence alignment.
        In both cases, acquiring this data requires time-consuming steps.
        Furthermore, variant calling using a linear reference can introduce a reference bias.
    }
\end{table*}

The library crate provides an interface for working with prefix-free graphs, mainly an iterator of a suffix array in sublinear space and linear time.
The library can be used from within the Rust programming language as follows:

\begin{minted}{rust}
let pfg = PFG::load("pfg.gfa");

for (i, (sa_i, id_i, pos_i)) in pfg.iter().enumerate() {
    println!("{}\t{}\t{}\t{}", i, sa_i, id_i, pos_i);
}
\end{minted}

In addition to the suffix array value, the iterator provides the segment ID and position values.
These are useful for the computation of several related data structures.
Similarly to \citet{2019boucher}, we can use them to compute the Burrows-Wheeler transform $\texttt{BWT}[i]$ by storing the preceding characters of each segment and then reporting the character at position $\texttt{pos}[i]-1$ of a segment $\texttt{id}[i]$.
This computation allows for space-efficient construction of r-index \cite{gagie2020fully} similar to MONI implementation \cite{rossi2022moni}, space-efficient construction of Wheeler graphs similar in nature to the implementation in \citet{2022pfwg}, and with the use of predecessor queries, space-efficient construction of a tag array introduced in \citet{2022maria}.

\section{Conclusion}
We introduced the prefix-free graph as a standalone data structure in pangenomics and showed how to use it to build a suffix array of a pangenome in sublinear space.
There are several possible directions for future research on prefix-free graphs.
The first direction may be to answer the question of which additional stringological algorithms and data structures can use prefix-free graphs to improve their applicability to such vast and repetitive datasets as pangenomic data.
Next, the choice of trigger words is, so far, mostly undiscovered territory.
From a computational perspective, the set of trigger words minimizing the size of a resulting prefix-free graph is desirable since it would allow the analysis of larger datasets.
Furthermore, trigger words offer great flexibility in the construction process.
We hypothesize this flexibility could be used for the construction of prefix-free graphs based on biologically significant strings.
In our experiments, we used the stop codon sequences as the trigger words, but other motives, such as different binding sites, recombination hotspots, or repetitive elements, are possible and could be used to capture particular biological phenomena.

\section{Online Resources}
The sources for the prefix-free graphs are available via
\begin{itemize}
\item \href{https://github.com/andynet/pfg}{https://github.com/andynet/pfg}.
\end{itemize}

\begin{acknowledgments}
This research was funded by a grant from the European Union’s Horizon 2020 research and innovation programme under the Marie Skłodowska-Curie grant agreement No 956229 (ALPACA) and by a grant from Slovak Research Grant Agency VEGA 1/0538/22.
\end{acknowledgments}

\bibliography{sources}

\end{document}